\input harvmac
\input epsf
\noblackbox
\newcount\figno
\figno=0
\def\fig#1#2#3{
\par\begingroup\parindent=0pt\leftskip=1cm\rightskip=1cm\parindent=0pt
\baselineskip=11pt
\global\advance\figno by 1
\midinsert
\epsfxsize=#3
\centerline{\epsfbox{#2}}
\vskip 12pt
\centerline{{\bf Figure \the\figno} #1}\par
\endinsert\endgroup\par}
\def\figlabel#1{\xdef#1{\the\figno}}
\def\pano{\par\noindent}
\def\smno{\smallskip\noindent}
\def\meno{\medskip\noindent}

\font\cmss=cmss10
\font\cmsss=cmss10 at 7pt
\def\rlx{\relax\leavevmode}
\def\inbar{\vrule height1.5ex width.4pt depth0pt}
\def\IC{\relax\,\hbox{$\inbar\kern-.3em{\rm C}$}}
\def\IR{\relax{\rm I\kern-.18em R}}
\def\IN{\relax{\rm I\kern-.18em N}}
\def\IP{\relax{\rm I\kern-.18em P}}
\def\ZZ{\rlx\leavevmode\ifmmode\mathchoice{\hbox{\cmss Z\kern-.4em Z}}
 {\hbox{\cmss Z\kern-.4em Z}}{\lower.9pt\hbox{\cmsss Z\kern-.36em Z}}
 {\lower1.2pt\hbox{\cmsss Z\kern-.36em Z}}\else{\cmss Z\kern-.4em Z}\fi}
\def\narrowplus{\kern -.04truein + \kern -.03truein}
\def\narrowminus{- \kern -.04truein}
\def\narrowminussub{\kern -.02truein - \kern -.01truein}

\def\cl{\centerline}

\def\o#1{\overline{#1}}

\def\ra{\rangle}

\def\type{type$\,{\rm II}$}

\def\typea{type$\,{\rm II}\,{\rm A}$}
\def\typeb{type$\,{\rm II}\,{\rm B}$}
\def\Typeb{Type$\,{\rm II}\,{\rm B}\ $}

\def\sqr#1#2{{\vcenter{\vbox{\hrule height.#2pt
 \hbox{\vrule width.#2pt height#1pt \kern#1pt
 \vrule width.#2pt}\hrule height.#2pt}}}}
\def\square
 {\mathop{\mathchoice{\sqr{12}{15}}{\sqr{9}{12}}{\sqr{6.3}{9}}{\sqr{4.5}{9}}}}


\lref\rKKS{S. Kachru, J. Kumar and E. Silverstein, {\it Vacuum Energy 
  Cancellation in a Nonsupersymmetric Strings}, hep-th/9807076 }

\lref\rKSb{S. Kachru and E. Silverstein,
{\it On Vanishing Two Loop Cosmological Constant in Nonsupersymmetric Strings},
 hep-th/9810129} 

\lref\rJH{J.H. Harvey, {\it String Duality and Nonsupersymmetric
  Strings}, Phys. Rev. {\bf D59} (1999) 026002, hep-th/9807213}

\lref\rKS{S. Kachru and E. Silverstein, {\it Self-Dual Nonsupersymmetric 
Type II String Compactifications}, J.High Energy Phys. 9811 (1998) 001,
hep-th/9808056}

\lref\rshiutyea{G. Shiu and S.-H.H Tye, {\it Bose-Fermi Degeneracy and 
Duality in Non-Supersymmetric Strings}, hep-th/9808095}

\lref\rDS{M. Dine and E. Silverstein, {\it New M-Theory Backgrounds with
 Frozen Moduli}, hep-th/9712166}

\lref\rscherk{J. Scherk and J.H. Schwarz, {\it Spontaneous Breaking of
Supersymmetry through Dimensional Reduction},
Phys. Lett. {\bf B82} (1979) 60}

\lref\ranto{I. Antoniadis, E. Dudas and A. Sagnotti, {\it
 Supersymmetry Breaking, Open Strings and M-Theory}, hep-th/9807011 \semi
I. Antoniadis, G. D'Appollonio, E. Dudas and A. Sagnotti, {\it
 Partial Breaking of Supersymmetry, Open Strings and M-Theory}, hep-th/9812118}

\lref\rvw{C. Vafa and E. Witten, {\it Dual Pairs with N=1 and N=2 Supersymmetry
   in Four Dimensions}, Nucl. Phys. Proc. Suppl. {\bf 46} (1996) 225, 
   hep-th/9508064}

\lref\rkakutye{Z. Kakushadze and S.-H.H Tye, {\it Brane World}, hep-th/9809147}

\lref\rshiutyeb{
G.\ Shiu and S.-H.H.\ Tye, {\it TeV Scale Superstring and Extra Dimensions},
Phys.Rev. {\bf D58} (1998) 106007, hep-th/9805157}

\lref\rgimpol{
E.G.\ Gimon and J.\ Polchinski, {\it Consistency Conditions
for Orientifolds and D-Manifolds}, Phys.\ Rev.\ {\bf D54} (1996) 1667,
hep-th/9601038}

\lref\ranom{M. Berkooz, R.G. Leight, J. Polchinsky, N. Seiberg,
J.H. Schwarz, E. Witten, {\it 
Anomalies, Dualities, and Topology of D=6 N=1 Superstring Vacua},
Nucl.Phys. {\bf B475} (1996) 115, hep-th/9605184}

\lref\rsen{A. Sen, {\it BPS D-branes on Non-supersymmetric Cycles},
hep-th/9812031}

\lref\rfsen{A. Sen, {\it A Non-perturbative Description of the 
Gimon-Polchinski Orientifold}, Nucl.Phys. {\bf B489} (1997) 139, 
  hep-th/9611186 \semi
  {\it F-theory and the Gimon-Polchinski Orientifold}, 
   Nucl.Phys. {\bf B498} (1997) 135, hep-th/9702061 }

\lref\rbackb{Z. Kakushadze, G. Shiu and S.-H.H Tye, {\it \Typeb Orientifolds
 with NS-NS Antisymmetric Tensor Backgrounds}, 
Phys. Rev. {\bf D58} (1998) 086001, hep-th/9803141 \semi
  M. Bianchi, G. Pradisi and A. Sagnotti, {\it Toroidal Compactification
 and Symmetry Breaking in Open-String Theories}, 
    Nucl.Phys. {\bf B376} (1992) 365}

\Title{\vbox{\hbox{hep--th/9812158}
 \hbox{HUB--EP--98/76}}}
{\vbox{ \hbox{Orientifolds of Non-Supersymmetric,} 
\vskip 0.5cm
\hbox{\phantom{eeeeee} Asymmetric Orbifolds}}}
\smallskip
\centerline{Ralph Blumenhagen${}^1$ and Lars G\"orlich${}^2$}
\bigskip
\centerline{\it Humboldt-Universit\"at Berlin, Institut f\"ur 
Physik,}
\centerline{\it  Invalidenstrasse 110, 10115 Berlin, Germany }

\smallskip
\bigskip
\bigskip\bigskip
\centerline{\bf Abstract}
\noindent
We consider certain four dimensional supersymmetric and 
non-supersymmetric asymmetric orbifolds 
with vanishing cosmological constant up to two loops and gauge 
the world sheet parity transformation. 
This leads to new string vacua, in which
Dp and D(p-4) branes or 
Dp and $\o{\rm D}$(p-4) are identified. Moreover, it is shown that 
different degrees of supersymmetry can be realized in the bulk and on the 
brane. We show that for non-supersymmetric models the cosmological constant 
still vanishes at one loop order.

\footnote{}
{\pano
${}^1$ e--mail:\ blumenha@physik.hu-berlin.de
\pano
${}^2$ e--mail:\ goerlich@physik.hu-berlin.de
\pano}
\Date{12/98}
\newsec{Introduction}

In the recent past we have seen some interesting developments in the field 
of non-supersymmetric string theory. Motivated by AdS-CFT duality,
S. Kachru, J. Kumar and E. Silverstein (KKS) proposed a peculiar 
$\ZZ_2\times \ZZ_2$ asymmetric orbifold
of \typeb\ which breaks all supersymmetry but nevertheless features
a vanishing cosmological constant both at one and two loops in string 
perturbation theory \refs{\rKKS{--}\rKSb}.
Strong-weak duality of such models suggests that such a behaviour
may even persist at higher loops and non-perturbatively \rKS. For a slightly 
different model, by using duality to heterotic strings it was shown that there
exist non-zero  non-perturbative contributions to the
cosmological constant \rJH . Generalisations of the original bosonic
models to backgrounds described by free world sheet fermions were 
discussed in \rshiutyea. \pano
All the models studied so far contain  at best abelian gauge symmetries in the
tree level massless spectrum. In this paper we will pursue the question 
whether
one can construct four dimensional non-supersymmetric string vacua with 
vanishing one loop cosmological constant and non-abelian gauge symmetries in 
the massless spectrum.
It seems to be impossible to obtain perturbative heterotic models with
the desired features. In a simple example, assume a $\ZZ_2$ symmetry $f$ 
breaks one half of the
supersymmetry and another $\ZZ_2$ breaks the other half. Then, since all
supersymmetry comes from the right moving sector in the heterotic string, the
twisted sector $fg$ breaks already all supersymmetry, so that the argument
of KKS fails. Another kind of models with non-abelian gauge symmetries
are orientifolds which we will consider in this paper. \pano
Naively, it seems to be nonsense to gauge the world sheet parity operation 
for an asymmetric orbifold. It seems simply not to be a symmetry. However, 
as also suggested in \rkakutye\ one
can arrange a situation where, for instance,  the two asymmetric $\ZZ_2$
symmetries get 
exchanged by parity reversal. We will study examples of exactly this type, 
for which the entire discrete symmetry group is non-abelian. In the usual way
tadpole cancellation requires  the introduction of  open strings in the theory
which are allowed to end on certain D-branes. Depending on what exactly
the two $\ZZ_2$ symmetries are, we are  lead to introduce Dp and D(p-4) 
branes or Dp and $\o{\rm D}$(p-4) in the theory. We consider examples with 
either $p=9$ or $p=8$. 
As already proposed in \rDS, the two types of branes are transformed into 
each other by the asymmetric
$\ZZ_2$ actions in the open string sector. As a result one gets sort of a 
bound  state between these two different kinds of branes.\pano
By analysing the massless spectrum, we find that independent of the degree
of supersymmetry in the bulk, one finds $N=2$ supersymmetry in the
open string sector.
This is similar to the supersymmetry breaking mechanism first introduced
by Scherk and Schwarz \rscherk\ and recently discussed in the string framework
by various authors \ranto. 
This paper is organised as follows. In section 2 we provide some background 
material for the definition of the three types of models
we will consider in the remaining part of the paper. In section 3 we 
compute all non-oriented one-loop diagrams, showing that they are still
zero. Consequently one has the same number of bosonic and fermionic degrees
of freedom at every mass level. 
In section 4 we compute the massless spectra in the open string
sector and in section 5 we will end with some conclusions. 

\newsec{Asymmetric orbifolds}
In this section we introduce six asymmetric orbifolds with either $N=2$ or
$N=0$ supersymmetry in the bulk. However, let us start with some 
preliminaries.
T-duality exchanges \typeb\ (\typea) on a circle of radius $R$ with \typea\ 
(\typeb) on a circle of radius $\alpha'/R$. If one considers an even number
of circles at the self-dual radius $R=\sqrt{\alpha'}$, T-duality becomes 
a symmetry of the model and can be gauged. In all cases discussed in the 
following we will fix at least four
circles at the self-dual radius. Thus we start with a compactification
on the $SU(2)^4$ torus. 
Furthermore, from now on we set $\alpha'=1/2$.
It is known that T-duality can be understood as a left moving 
reflection and is therefore  denoted as $(-1,1)$.   
However, modding out $f=(-1,1)^4$ alone is not possible, for it does not
satisfy level matching. This can be seen by computing the partition 
function in the  ${\scriptstyle{f}}\square\limits_1$ sector 
\eqn\twist{ {\scriptstyle{f}}\square\limits_1 \sim 
      \left(\o\Theta_{0,1}(\o\tau)\right)^4 ,}
where $\Theta_{m,k}$ denote the $SU(2)_k$ theta-functions.
The 
${\scriptstyle{1}}\square\limits_f$ sector is determined by a modular 
S-transformation and is proportional to 
$\left(\o\Theta_{0,1}+\o\Theta_{1,1}\right)^4$.
Apparently this contains levels $\o h\in\ZZ+1/4$ which violate the 
level matching condition.  
As was observed in \rKKS\ this can be repaired by equipping $f$ with further 
shifts. In the following 
we will work with the $\ZZ_2$ shifts $A_1,A_2,A_3$ introduced in \rvw. 
Table 2 shows how the three kinds of shifts act on Kaluza-Klein
(KK) and winding states $(m,n)$ and to what left-right shift they 
correspond to.   
\smno
\cl{\vbox{
\hbox{\vbox{\offinterlineskip
\def\tablespace{height2pt&\omit&&\omit&&
 \omit&\cr}
\def\tablerule{\tablespace\noalign{\hrule}\tablespace}

\hrule\halign{&\vrule#&\strut\hskip0.2cm\hfil#\hfill\hskip0.2cm\cr
\tablespace
& shift && action on $(m,n)$  && $(X_L,X_R)$\ shift     &\cr
\tablerule
& $A_1$ && $(-1)^n$ && $\left({1\over 4R},-{1\over 4R}\right)$ &\cr
\tablespace
& $A_2$ && $(-1)^{m+n}$ && $\left({1\over 4R}+{R\over 2},-{1\over 4R}+
         {R\over 2}\right)$ &\cr
\tablespace
& $A_3$ && $(-1)^m$ && $\left({R\over 2},{R\over 2}\right)$ &\cr
\tablespace}\hrule}}}}
\cl{
\hbox{{\bf Table 1:}{\it ~~ $\ZZ_2$ shifts }}}
\smno
Note, that only $A_3$ is a left-right symmetric shift and thus can really 
be interpreted as an ordinary shift of the coordinate $X=X_L+X_R$. 
The existence of the asymmetric shift $A_1$ is a purely stringy effect and
can be understood as a shift in momentum space. In particular, in the open
string sector Wilson-lines get affected by these shifts.  
Nevertheless, the
partition function obtained after modding out each one  of the three shifts is
left-right symmetric reflecting the transformation property 
$\Omega A_i\Omega^{-1}=A_i$.
Under $T$-duality however, $A_1$ and $A_3$ get exchanged whereas $A_2$ is
invariant, $T A_1 T^{-1}=A_3$ and $TA_2 T^{-1}=A_2$.\pano
Due to this non-trivial transformation property, the discrete symmetries
$f=(-1,1;A_3)^4$ and $f=(1,-1;A_3)^4$ satisfy $f^2=A_2^4$ and are no 
longer $Z_2$ transformations. In order to make them of order two, we first
have to divide the $SU(2)^4$ torus by the asymmetric shift $A_2^4$. 
The resulting partition function is nothing else than the partition function
for the $SO(8)$ torus. Now, we could follow two possible paths. First,
we could realize the $SO(8)$ torus directly with some background metric
$G_{ij}$ and some non-vanishing background two form $B_{ij}$. Second,
we could continue to consider the $SO(8)$ torus as an orbifold of the
$SU(2)^4$ torus. In the following, we will follow the second path, but
keep in mind that our result has to be consistent with what one expects
to get following  the first path. In particular, in the open string sector
it is known that the background $B_{ij}$ field reduces the rank of the
gauge symmetries by a factor $2^{b\over 2}$, where $b=rk(B)$ \rbackb.\pano
The partition function in the  
${\scriptstyle{f}}\widetilde{\square\limits_1}$ sector 
is
\eqn\twist{\eqalign{ {\scriptstyle{f\,}}\widetilde{\square\limits_1} &=
      {\scriptstyle{f}}\square\limits_1 + 
      {\scriptstyle{fA_2^4}}\square\limits_1 +
      {\scriptstyle{f}}\square\limits_{A_2^4} +
      {\scriptstyle{fA_2^4}}\square\limits_{A_2^4}\cr
      &\sim 
      \left(\o\Theta_{0,4}-\o\Theta_{4,4}\right)^4 +
      \left(\o\Theta_{-2,4}-\o\Theta_{2,4}\right)^4.\cr}}
The second term vanishes, so that after a modular S-transformation
on obtains 
\eqn\afts{ {\scriptstyle{1\,}}\widetilde{\square\limits_f} \sim 
    \left(\o\Theta_{-3,4}+\o\Theta_{-1,4}+
        \o\Theta_{1,4}+\o\Theta_{3,4} \right)^4,}
which has conformal weight
$\o h=1/4$ violating level matching. This can be repaired by
compactifying on a further circle with shift $A_2$.\pano
Having defined various symmetry action for toroidal compactifications, 
we can list the six models we will study in the course of this paper
\meno
\cl{\vbox{
\hbox{\vbox{\offinterlineskip
\def\tablespace{height2pt&\omit&&\omit&&\omit&&\omit&&
 \omit&\cr}
\def\tablerule{\tablespace\noalign{\hrule}\tablespace}

\hrule\halign{&\vrule#&\strut\hskip0.2cm\hfil#\hfill\hskip0.2cm\cr
\tablespace
& && $f$ && $g$   && $\Omega'$ && susy   &\cr
\tablerule
& Ia && $(-1,1;A_3)^4\, A_2\, A_3$ && $(1,-1;A_3)^4\, A_2\, A_3$ && 
$\Omega$ && N=2 &\cr
\tablespace
& Ib && $(-1,1;A_3)^4\, A_2\, A_3\ (-1)^F$ && $(1,-1;A_3)^4\, A_2\, A_3\ 
(-1)^F$ && $\Omega$ && N=0 &\cr
\tablerule
& IIa && $(-1,1;A_3)^4\, A_2\, A_3\ (-1)^{F_L}$ && $(1,-1;A_3)^4\, A_2\, A_3 \
(-1)^{F_R}$ && $\Omega$ && N=2 &\cr
\tablespace
& IIb && $(-1,1;A_3)^4\, A_2\, A_3\ (-1)^{F_R}$ && $(1,-1;A_3)^4\, A_2\, A_3\ 
(-1)^{F_L}$ && $\Omega$ && N=0 &\cr
\tablerule
& IIIa && $(-1,1;A_3)^4\, A_2\, A_3\ (-1)^{F_L}$ && $(1,-1;A_3)^4\, A_3\, A_2 \
(-1)^{F_R}$ && $\Omega\,P$ && N=2 &\cr
\tablespace
& IIIb && $(-1,1;A_3)^4\, A_2\, A_3\ (-1)^{F_R}$ && $(1,-1;A_3)^4\, A_3\, A_2\ 
(-1)^{F_L}$ && $\Omega\,P$ && N=0&\cr
\tablespace}\hrule}}}}
\cl{
\hbox{{\bf Table 2:}{\it ~~ Definition of models }}}
\meno
Apparently, for model III the world sheet parity reversal $\Omega$ is not a 
symmetry, but combining it with a permutation $P$ of the $X_5$ and $X_6$ 
directions one obtains a symmetry $\Omega'=\Omega\,P$.
However, one has to be a bit more careful. For  $\Omega$ to be a symmetry of 
the models I and II we had to start with \typeb . Since the permutation $P$
changes  a chiral ten dimensional spinor into an antichiral spinor, we should 
better start with \typea\ to guarantee that $\Omega\, P$ is indeed a 
symmetry. \pano
The first four models in Table 2 are similar to those considered by J. Harvey 
in \rJH,
whereas the last two ones are similar to the original one of KKS 
\footnote{$^1$}{As in reference \rJH\  we consider f and g as commuting 
generators in the point group with the sector 
${\scriptstyle{f}}\square\limits_g=0$.
This implies $Z={1\over 2}(Z_f+Z_g+Z_{fg}-Z_1)$ where every term preserves
some supersymmetry.}\rKKS. 
Before gauging $\Omega'$ all six models have $32$ massless bosonic degrees
of freedom and $32$ massless fermionic degrees of freedom in the untwisted 
sector. Models I and II also contain $64$ bosonic and fermionic degrees
of freedom in the $R$ twisted sector. Since in model III $R=f\,g\,A_2^4$ 
still contains 
a shift in the $X_5$ and $X_6$ directions, there do not appear additional
massless states in the $R$ twisted sector.  
Furthermore, the complete perturbative massive spectrum is also bose-fermi
degenerated, leading to a vanishing cosmological constant at one loop.
Now, we would like to gauge also the parity reversal $\Omega'$. 
Considering only the massless spectrum one finds that the degrees of freedom 
arising in the closed string sector are exactly halved by this projection. 
However, one expects to find massless tadpoles arising in the Klein bottle
amplitude, which makes it necessary to introduce an open string sector in the 
theory. One might argue that the open string only sees the left-right 
symmetric part of the orbifold and thus should live in a world with $N=2$
supersymmetry independently of the supersymmetry in the bulk.
\newsec{Tadpole cancellation}
Computationally,  the new aspect we are facing is to gauge a non-abelian 
discrete symmetry group containing $\Omega'$. 
Using the relation
$\Omega'\, f\, \Omega'=g$ and that $f$ and $g$ commutes 
one can easily show that the two symmetries
\eqn\dvier{ \theta=\Omega' f\quad\quad {\rm and}\quad\quad r=f }
generate the non-abelian group $D_4$ \footnote{$^1$}{The non-abelian discrete 
symmetry group $D_4$ has eight elements and contains two generators $r$ and
$\theta$ satisfying the relations $r^2=1$, 
$\theta^4=1$ and $r\theta=\theta^3 r$. Moreover, $D_4$ has 5 irreducible 
representations,
four of them are one dimensional and one is two dimensional}.
Let us give a formal argument
how the partition function looks like. 
Modding out a closed string by $D_4$ one knows that, since one has to sum 
only 
over commuting twists along the two fundamental cycles the partition function
can be written as a sum over abelian orbifolds
\eqn\part{ Z={1\over 2} \left( Z_{\theta}+Z_{r}+Z_{r\theta}-Z_{\theta^2}
  \right) .}
$Z_{\theta}$ is a $\ZZ_4$ orbifold with in our case elements 
$(1,\Omega'f,R,\Omega'g)$,
$Z_{r}$ is a $\ZZ_2\times\ZZ_2$ orbifold with elements $(1,f,g,R)$, 
$Z_{r\theta}$ is also a 
$\ZZ_2\times\ZZ_2$ orbifold with elements  $(1,R,\Omega',\Omega'R)$ and 
finally 
$Z_{\theta^2}$ is a $\ZZ_2$ orbifold with elements $(1,R)$. Here, we have 
formally identified sectors twisted by $\Omega'$ as open string sectors.
Note, that $Z_{\theta}$ and $Z_{\theta^2}$ do only contain 
closed string sectors.
Moreover, due to the shifts in $f$  one convinces oneself that 
the orientifold $Z_{\theta}$ does not
lead to any massless tadpole in the tree channel Klein bottle amplitude. 
Therefore, there is no need to introduce open strings here. In other words,
would we not had included the twists in $f$ and $g$, we would now had been
in trouble with explaining what a sector twisted by $\Omega'f$ should be.\pano
Summarising, the only open string sector appears in the abelian
orientifold  ${1\over 2}Z_{r\theta}$, which is obviously related to
the kind of orientifold firstly studied by Gimon and Polchinski (GP) \rgimpol.
This simple argument shows that in the open string sector
the degrees of freedom in our non-abelian orientifold model are expected to be
half the number of degrees of freedom in the related GP model. This 
immediately implies that the one-loop cosmological constant in the open
string sector also vanishes. In the following we will see, that due to the
shifts in the 5-6 directions the mass levels of the $95$ and $59$
strings get shifted against the mass levels of the $99$ and $55$
open strings. But still the one-loop cosmological constant vanishes.\pano 
The GP model contains 9-branes and 5-branes which are now further related by 
the action of $f$. $f$ contains a T-duality transformation
in four directions, thus simply exchanging the two kinds of branes. 
Furthermore, positions of 5-branes in the transverse directions get mapped 
onto Wilson lines of 9-branes around those four circles. Moreover,
due to the $A_2$ shift in the $x_5$ direction, the Wilson lines
inside the 9  and 5-branes have to be shifted
against each other. In a T-dual version the $x_5$ positions of the
8-branes and 4-branes would have to be on opposite sides of the
circle. This immediately implies that there are no massless modes from 
open strings stretched between two such branes.
In the next
part we will compute in detail the remaining three different one loop
contributions to the cosmological constant, namely the Klein bottle, cylinder 
and M\"obius strip amplitude. 

\subsec{Klein bottle}

In the loop channel Klein bottle amplitude one has to sum over all sectors
\eqn\kbtwist{   K(t)_{g,h}={\rm Tr}_{g}\left( \Omega\, h\, 
            e^{-2\pi t(L_0+\o{L}_0)}\ P\ S\ \right) ,}
for which $g$ and $\Omega h$ commute. 
This is in agreement with the consistency conditions derived in \rgimpol\
for the tree channel Klein bottle amplitude. 
$P$ denotes the GSO projection and $S={1\over 2}(1+A_2^4)$ denotes
the projection onto states invariant under the asymmetric shift $A_2^4$. 
Applied to our case \kbtwist\  leads to the following contribution to the 
cosmological constant 
\eqn\kbtot{ \Lambda_{KB}\sim 4\int_0^\infty {dt\over t^3} 
    {1\over 4} {\rm Tr}_{1,R}\left( \Omega\, (1+f+g+R)\, 
    e^{-2\pi t(L_0+\o{L}_0)}\ P\, S\ \right),}
where one only has to take the trace over the untwisted and $R=fg A_2^4$ 
twisted sectors. Moreover, since $\Omega$ exchanges left and right movers the 
trace needs only to be taken over the NS-NS and R-R sectors. Therefore
the space time fermion number operator $(-1)^F$ appearing in the definition of
 $f$ and $g$ does not matter in the Klein bottle amplitude.
The effect of the $(-1)^{F_L,F_R}$ insertions in $f$ and $g$ is also 
marginal, for it only changes some of the signs in front of terms containing 
contributions
from the spin structure ${\scriptstyle{+}}\square\limits_+$
which is zero anyway. Thus, also for the non-supersymmetric cases Ib and IIb 
the Klein bottle amplitude vanishes. Moreover, the shifts contained in $f$ 
and $g$ permute the $16$
different fixed points of $R$ and therefore the
contributions $K(t)_{R,f}$ and $K(t)_{R,g}$ are identical to zero.\pano
Summarising, the Klein bottle amplitude for model I and II can be written as
\eqn\kbres{\eqalign{   \Lambda^{I,II}_{KB} \sim (1-1)\, &4\int_0^\infty 
      {dt\over t^3}\Biggl\{
      {f_4^8(e^{-2\pi t})\over f_1^8(e^{-2\pi t})}
            \left(\sum_{m\in\ZZ} e^{-\pi t{m^2\over \rho}} \right)^2\
{1\over 2}\biggl[ \left(\sum_{m\in\ZZ} e^{-\pi t{m^2\over \rho}} \right)^4+\cr
        &\left(\sum_{m\in\ZZ} (-1)^m\, e^{-\pi t{m^2\over \rho}} \right)^4+
           \left(\sum_{n\in\ZZ} e^{-\pi t{\rho n^2}} \right)^4 +
     \left(\sum_{n\in\ZZ} (-1)^n \, e^{-\pi t{\rho n^2}} \right)^4 \biggr] \cr
          & +8 {f_4^4(e^{-2\pi t})f_3^4(e^{-2\pi t})\over 
             f_1^4(e^{-2\pi t})f_2^4(e^{-2\pi t})}
            \left(\sum_{m\in\ZZ} (-1)^m e^{-\pi t{m^2\over \rho}} \right)^2 
        \Biggr\}    \cr} .}
Here we have defined $\rho=r^2/\alpha'$ which is actually one at the 
self-dual radius.
However, we keep it in the formulae in order to follow the different volume
factors in the amplitudes. We have also set the radii of the 5-6 circles
to the self-dual radius. 
In order to detect massless tadpoles one has to transform the amplitude 
into the tree channel by a modular transformation $t={1\over 4l}$.
Since the Poisson resummation formula implies
\eqn\pois{ \sum_m (-1)^m\, e^{-\pi {m^2\over \rho}{1\over 4l}}=
   \sum_m e^{-4\pi\rho l \left(m+{1\over 2}\right)^2} }
there are no new tadpoles arising from the second term, the $K(t)_{1,f}$ and
$K(t)_{1,g}$ sectors, in \kbres.
The only tadpoles have their origin in the first
term of \kbres, which up to factor $ {1\over 2}$ yields exactly the same 
tadpole as in the associated GP model. \pano
The computation for model III is slightly different. Let us, for instance,
determine how $\Omega P$ acts on the KK and winding states in the 5-6
directions:
\eqn\opact{\Omega P\vert m_5,n_5;m_6,n_6\ra = \vert m_6,-n_6;m_5,-n_5\ra, }
so that  only states with $m_5=m_6$ and $n_5=n_6$ contribute in the trace.
The complete calculation yields the following amplitude
\eqn\kbdrei{\eqalign{   \Lambda^{III}_{KB} \sim (1-1)\, &4\int_0^\infty 
      {dt\over t^3} \Biggl\{
       {f_4^8(e^{-2\pi t})\over f_1^8(e^{-2\pi t})}
            \left(\sum_{m\in\ZZ} e^{-2\pi t{m^2\over \rho}} \right)
            \left(\sum_{n\in\ZZ} e^{-2\pi t{\rho n^2}} \right) \ 
        {1\over 2}\biggl[ \left(\sum_{m\in\ZZ} 
             e^{-\pi t{m^2\over \rho}} \right)^4+ \cr
             &\left(\sum_{m\in\ZZ} (-1)^m
             e^{-\pi t{m^2\over \rho}} \right)^4+
             \left(\sum_{n\in\ZZ} e^{-\pi t{\rho n^2}} \right)^4 +
       \left(\sum_{n\in\ZZ} (-1)^n e^{-\pi t{\rho n^2}} \right)^4 \biggr] +\cr
          & 8 {f_4^4(e^{-2\pi t})f_3^4(e^{-2\pi t})\over 
             f_1^4(e^{-2\pi t})f_2^4(e^{-2\pi t})}
                 \left(\sum_{m\in\ZZ} e^{-2\pi t{m^2\over \rho}} \right)
                 \left(\sum_{n\in\ZZ} (-1)^n e^{-2\pi t{\rho n^2}} \right)
      \Biggr\} \cr} }
The second term in \kbdrei\ does not generate a tadpole either and the
two tadpoles from the first term scale with volume factors $V_9$ and $V_5$,
pointing already to compensating brane contributions from D8 and D4 branes.
The next step is to introduce an open string sector, on which one first has
to determine the action of the asymmetric generators $f$ and $g$.  

\subsec{Cylinder and M\"obius amplitude}

First, we would like to discuss the examples I and II. 
Since it is essentially the GP
computation we have to introduce D9-branes and D5($\o{\rm D}$5)-branes in the 
theory in order
to cancel the tadpoles from the Klein bottle amplitude.
We have to  arrange the branes in such a way that they respect the symmtries
$A_2^4,f,g$ we would like to mod out. 
For instance, since the shift $A_2^4$ acts on the Wilson lines of the 
D9-branes
and the positions of the D5($\o{\rm D}$5)-branes, respectively, we can not 
choose them equal for all D9- and D5($\o{\rm D}$5)-branes. At best, we can 
have half of the D9-branes
with Wilson lines $\Theta^i$ and the other half with Wilson-lines
$\Theta^i+{1\over 2R}$, so that they get mapped to each other by $A_2^4$.
Analogously, we can place half the D5($\o{\rm D}$5)-branes at a fixed point 
$X_i$ and the other half at the shifted fixed point $X_i+{R\over 2}$.
We will see, that this has exactly the effect of reducing the rank of the 
gauge group in the expected way.
In order not to confuse the reader to much, we restrict ourselves
to this case of maximal gauge symmetry.\pano 
Moreover, under T-duality $(-1,1)^4$ the D9- and D5- branes get exchanged, 
where positions $(X^i_\mu )_{i=1,\ldots,N_5}$
of D5($\o{\rm D}$5)-branes in the transverse space 
are mapped to Wilson lines $(\Theta^i_\mu )_{i=1,\ldots,N_9}$
of D9-branes around those four cycles. Since we would like to gauge $f$ we 
should guarantee that it is indeed a symmetry in the open string sector, 
as well. To this end, one needs the same number $N$ of D9- and 
D5($\o{\rm D}$5) branes
just from the very beginning. Furthermore, the branes need to carry 
Wilson lines or need to have transverse positions, respectively, in just the 
right way to get mapped onto each other by $f$. 
We also restrict ourselves to the case where all D9-brane Wilson
lines around cycles
in the 5-6 torus are equal. Of course, the  D5($\o{\rm D}$5)-brane Wilson 
lines  are shifted by one half but are also equal among themselves.  
In the original GP model, D9 brane Wilson line moduli and 
D5($\o{\rm D}$5)-brane
position moduli were completely independent, whereas now they become related.
It is in this sense that we speak of a bound state of 
D9 and D5($\o{\rm D}$5) branes.
Figure 1 shows that, when $f$ maps Wilson lines of an open string
$(\Theta^1,\Theta^2)$ to positions $(X^1,X^2)$ then $g$ maps it to 
the inverse positions $(-X^1,-X^2)$.
\fig{}{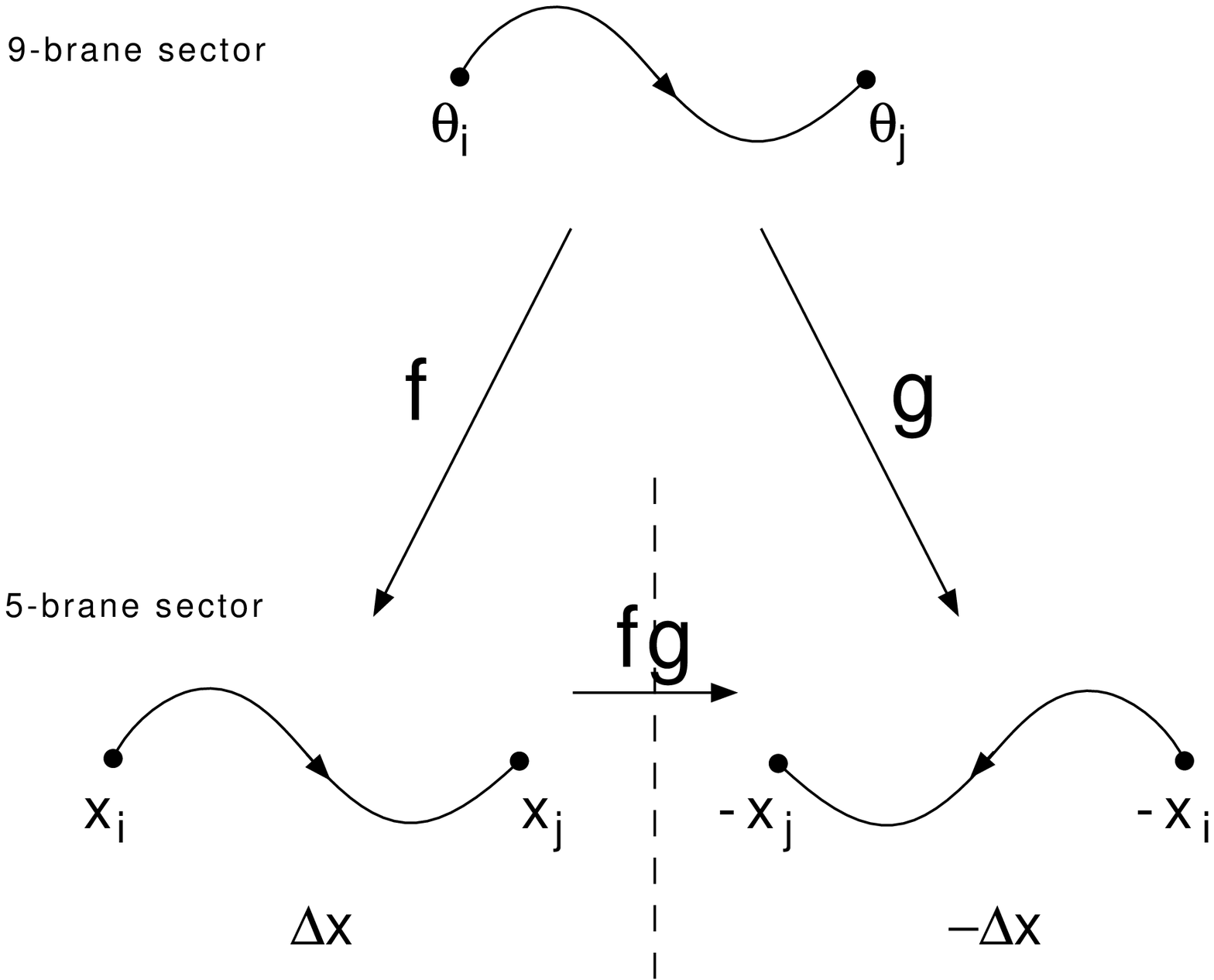}{7truecm}
\pano
As visualised in figure 2, taking into account $\Omega\delta p\Omega=
\delta p$ and $\Omega\delta x\Omega=-\delta x$,  it can be shown that
this is compatible with the relation $\Omega f\Omega=g$ now realized in the
open string sector. 
\fig{}{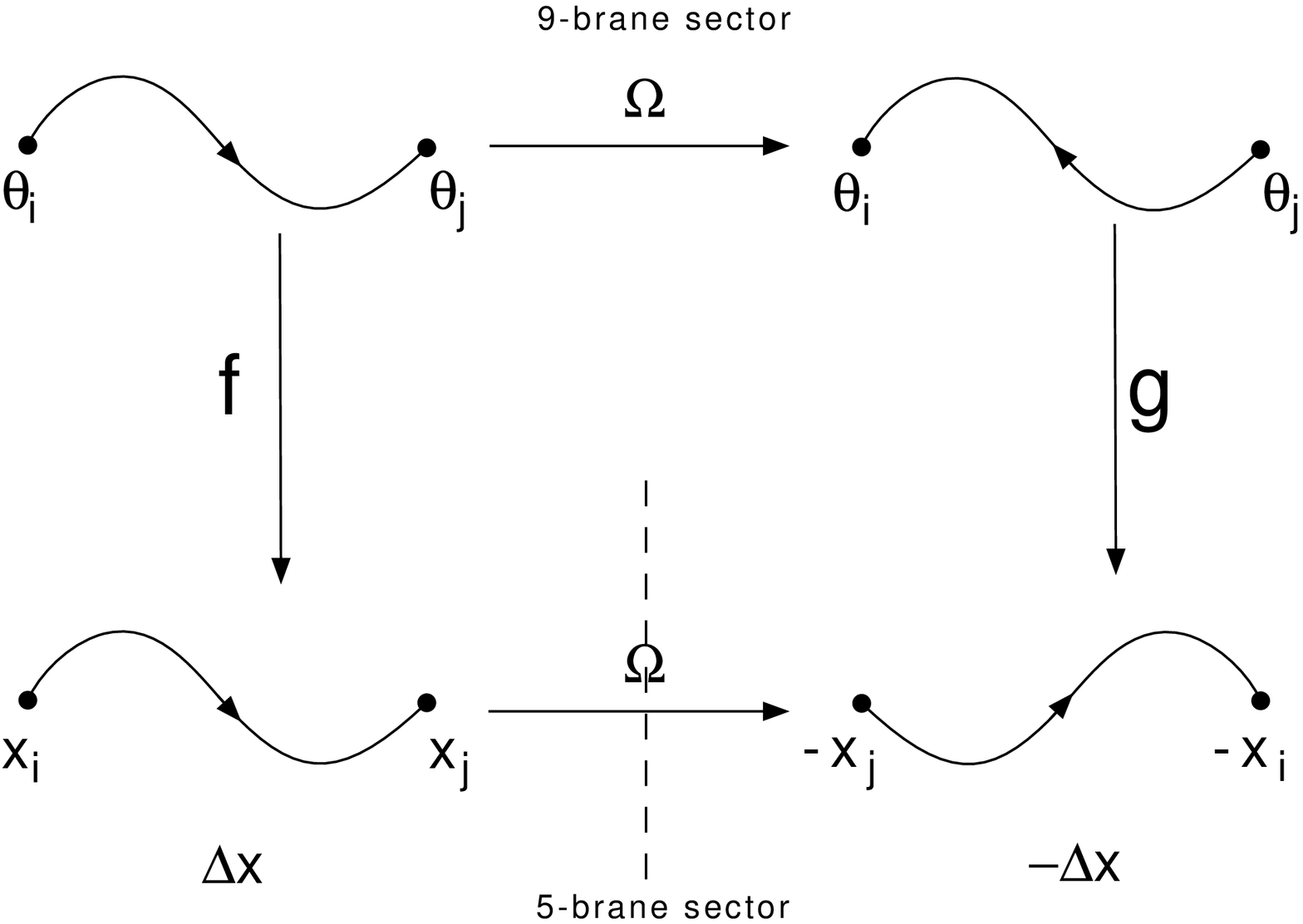}{7truecm}
\pano
Depending on the boundary conditions, the action of the various shifts in 
the open string sector is presented in Table 3.
\smno
\cl{\vbox{
\hbox{\vbox{\offinterlineskip
\def\tablespace{height2pt&\omit&&\omit&&
 \omit&\cr}
\def\tablerule{\tablespace\noalign{\hrule}\tablespace}

\hrule\halign{&\vrule#&\strut\hskip0.2cm\hfil#\hfill\hskip0.2cm\cr
\tablespace
& shift && NN   && DD     &\cr
\tablerule
& $A_1$ && $\Theta\to \Theta+{1\over 2R}$ && $(-1)^n$ &\cr
\tablespace
& $A_2$ && $(-1)^{m}$ && $(-1)^n$ &\cr
&       && $\Theta\to \Theta+{1\over 2R}$ && $X\to X+{R\over 2}$ &\cr
\tablespace
& $A_3$ && $(-1)^m$ && $X\to X+{R\over 2}$ &\cr
\tablespace}\hrule}}}}
\cl{
\hbox{{\bf Table 3:}{\it ~~ Action of shifts on open string}}}
\smno
The asymmetric fermion number operator
$(-1)^{F_L}$ flips the charges of all R-R fields, thus transforming a 
D-brane into its charge conjugate anti-brane  \rsen. 
Thus, the orientifolds in class I contain D9 and D5 branes whereas
the orientifolds in class II must have D9 and $\o{\rm D}$5 branes.
Note, that contrary to the examples of non-supersymmetric 
Dp-$\o{\rm D}$p systems discussed
recently in a series of papers, a pair of flat D9 and $\o{\rm D}$5 branes
does not break all the supersymmetry. 
We conclude that all the asymmetric operations contained in $f$ and $g$ have 
a well defined action in the open string sector.\pano 
To describe the action of $f$ on the Chan-Paton factors we introduce a matrix
$\gamma_f$ which must be of the form
\eqn\gammaf{  \gamma_f=\left(\matrix{  0  &  f_{95} \cr
                      f_{95}^{-1} & 0 \cr} \right), }
where $f_{95}$ is an $N\times N$ matrix. Remember, that the actions of all the
other symmetries like $\Omega$ and $R$ were block diagonal in the GP model. 
Given the choice made in GP
\eqn\gammao{\gamma_\Omega=\left(\matrix{  1  &  0 \cr
                           0 & M \cr}\right),\quad\quad 
     \gamma_{R}=\left(\matrix{  M  &  0 \cr
                           0 & M \cr}\right)\quad {\rm with}\quad 
     M=\left(\matrix{ 0   &  i \cr
                           -i & 0 \cr}\right), }
consistent with all conditions for the $\gamma$ matrices  we can choose $f_{95}=1$.
\pano
Generally, the cylinder amplitude is defined by
\eqn\cyl{\Lambda_{C}\sim \int_0^\infty {dt\over t^3}\, 
      {1\over 4}\, {\rm Tr}_{99,55,95,59}\left( (1+f+g+R)\, 
      e^{-2\pi t L_0}\ P\ S\ (-1)^F \right).}
Since $f$ and $g$ change the type of brane, the trace with $f$ and $g$ 
insertions is trivially zero.
We obtain for the complete cylinder amplitude 
\eqn\cylres{\Lambda^{I,II}_{C} \sim  \int_0^\infty {dt\over t^3}
       \left( Z_{99}+Z_{55}+Z_{95}+Z_{59} \right) }
with the contribution from the 99 open strings being
\eqn\cylnine{\eqalign{ Z_{99}=&{f_3^8(e^{-\pi t})-f_4^8(e^{-\pi t})-
              f_2^8(e^{-\pi t})\over f_1^8(e^{-\pi t})} 
          \left(\sum_{m\in\ZZ} e^{-2\pi t{m^2\over \rho}} 
        \right)^2 \cr  
         &{1\over 2}\biggl[ \sum_{i,j\in 9} (\gamma_{1,9})_{ii}
         (\gamma_{1,9})_{jj} \biggl\{
         \prod_{k=1}^4 \sum_{m\in\ZZ}
         e^{-2\pi t\left( {m\over r}+\Theta^i_k-\Theta^j_k\right)^2/
           \alpha'^{-1}}+ \cr
         &\prod_{k=1}^4 \sum_{m\in\ZZ} (-1)^m
         e^{-2\pi t\left( {m\over r}+\Theta^i_k-\Theta^j_k\right)^2/
           \alpha'^{-1}} \biggr\} \biggr] + \cr
          &4\, { f_3^4 f_4^4-f_4^4 f_3^4 \mp f_2^4 f_0^4\pm f_0^4 f_2^4 \over 
  f_1^4 f_2^4} \left(\sum_{m\in\ZZ} e^{-2\pi t{m^2\over \rho}} \right)^2
           \ \sum_{J=1}^{16} \left( Tr(\gamma_{R,J})\right)^2.\cr}}
For open strings stretched between two D5-branes we obtain         
\eqn\cylfive{\eqalign{ Z_{55}=&{f_3^8(e^{-\pi t})-f_4^8(e^{-\pi t})-
              f_2^8(e^{-\pi t})\over f_1^8(e^{-\pi t})} 
          \left(\sum_{m\in\ZZ} e^{-2\pi t{m^2\over \rho}} 
        \right)^2\cr  
      &{1\over 2}\biggl[ \sum_{i,j\in 5} (\gamma_{1,5})_{ii}
         (\gamma_{1,5})_{jj}\biggl\{ 
         \prod_{k=1}^4 \sum_{n\in\ZZ}
         e^{-2\pi t\left( {n r}+X^i_k-X^j_k\right)^2/\alpha'}+\cr
          &\prod_{k=1}^4 \sum_{n\in\ZZ} (-1)^n
         e^{-2\pi t\left({n r}+X^i_k-X^j_k\right)^2/
           \alpha'} \biggr\}\biggr] +\cr
          &4\, { f_3^4 f_4^4-f_4^4 f_3^4 \mp f_2^4 f_0^4\pm f_0^4 f_2^4 \over 
  f_1^4 f_2^4} \left(\sum_{m\in\ZZ} e^{-2\pi t{m^2\over \rho}} \right)^2
      \ \sum_{I=1}^{16} \left( Tr(\gamma_{R,I})\right)^2.\cr }}
Finally, the 95 and 59 sector together  contribute
\eqn\cylnf{\eqalign{ Z_{95}=&2\, { f_3^4 f_2^4\pm f_4^4 f_0^4- f_2^4 f_3^4\mp
        f_0^4 f_4^4 \over f_1^4  f_4^4 } 
       \left(\sum_{m\in\ZZ} e^{-2\pi t{m^2\over \rho}} \right)
       \left(\sum_{m\in\ZZ} e^{-2\pi t{(m+{1\over 2})^2\over \rho}} \right) 
      Tr(\gamma_{1,5}) Tr(\gamma_{1,9}) +\cr
        &2\, { f_3^4 f_0^4\pm  f_4^4 f_2^4\mp f_2^4 f_4^4- f_0^4 f_3^4 \over 
        f_1^4 f_3^4} \left(\sum_{m\in\ZZ} e^{-2\pi t{m^2\over \rho}} \right)
        \left(\sum_{m\in\ZZ} e^{-2\pi t{(m+{1\over 2})^2\over \rho}} \right)
        Tr(\gamma_{R,5}) Tr(\gamma_{R,9}), \cr } }
where $f_0=0$ denotes the trace for the spin structure 
${\scriptstyle{+}}\square\limits_+$\ .
The upper sign in the above equations belongs to model I and the lower one 
to model II. 
We have deliberately written the complete partition function, for we would 
like to discuss the physical relevance of the different signs. 
For model I everything is like in the GP case, we have computed the 
one loop graph for the interaction of two D-branes, which of course vanishes
due to supersymmetry or bose-fermi degeneracy, respectively. 
For model II two extra signs occur. One is due to the $(-1)^F$ action
in $R$ and
the other is due to the fact that we have to consider a $\o{\rm D}$5 brane
instead of a D5 brane. This is achieved in the loop channel by changing the 
GSO projection from ${1\over 2}(1+(-1)^f)$ to ${1\over 2}(1-(-1)^f)$ implying 
that the tree level
exchange of R-R fields contribute with the other sign.
This nicely confirms that we indeed need $\o{\rm D}$5 branes for model
II, since otherwise the brane would be attracted to a D9-brane and we would not
get a stable background. \pano
Finally, we compute the M\"obius amplitude 
\eqn\moebi{\Lambda_{M}\sim \int_0^\infty {dt\over t^3} 
      {1\over 4} {\rm Tr}_{99,55}\left(\Omega\, (1+f+g+R)\, e^{-2\pi t L_0}\ 
       P\ S\ (-1)^F \right)}
Again, since $f$ and $g$ change the type of brane the traces with $f,g$
insertions vanish. 
\eqn\moebires{\eqalign{   \Lambda^{I,II}_{M} \sim  (1-1)&\int_0^\infty 
      {dt\over t^3}-{f_2^8(i\,e^{-\pi t})\over 
         f_1^8(i\,e^{-\pi t})} \left(\sum_{m\in\ZZ} e^{-2\pi t{m^2\over \rho}} \right)^2 \cr
        {1\over 2}&\biggl[ Tr(\gamma_{\Omega,9}^{-1} \gamma_{\Omega,9}^{T})
          \left(\sum_{m\in\ZZ} e^{-2\pi t {m^2\over \rho}}\right)^4 +
          Tr(\gamma_{\Omega,9}^{-1} \gamma_{\Omega,9}^{T})
          \left(\sum_{m\in\ZZ} (-1)^m e^{-2\pi t {m^2\over \rho}}\right)^4 +
\cr
            & Tr(\gamma_{\Omega\, R,5}^{-1} \gamma_{\Omega\, R,5}^{T})
          \left(\sum_{n\in\ZZ} e^{-2\pi t \rho n^2}\right)^4 +
           Tr(\gamma_{\Omega\, R,5}^{-1} \gamma_{\Omega\, R,5}^{T})
       \left(\sum_{n\in\ZZ} (-1)^n e^{-2\pi t\rho n^2}\right)^4\biggr]. \cr }}
Here, the extra $(-1)^F$ factor in $R$ in model II means that the orientifold 
plane carries the opposite charge, so that we have to put a
$\o{\rm D}$5-brane in front of an $\o{\rm O}$5-plane in order to get 
zero force.
Working out the tadpole cancellation conditions one finds that they cancel
exactly for $32$ D9-branes and $32$ D5($\o{\rm D}$5)-branes. \pano
Before we discuss the massless spectra in the open string sector, we would
like to briefly describe what happens for model III.
Apparently, since we are in \typea\ we should better introduce odd dimensional
branes.  Indeed, the tree channel Klein bottle amplitude contains an exchange
of a R-R 9-form potential showing that we should have some D8 branes and
hence $\o{\rm D}$4-branes involved. Denote by $a$ and $b$ the two fundamental 
cycles of the torus spanning the 5-6 directions. Both the D8 and the 
$\o{\rm D}$4-brane
must wrap around the cycle $a+b$ which is invariant under $P$ and must
be localised on the antisymmetric cycle $a-b$.
\fig{}{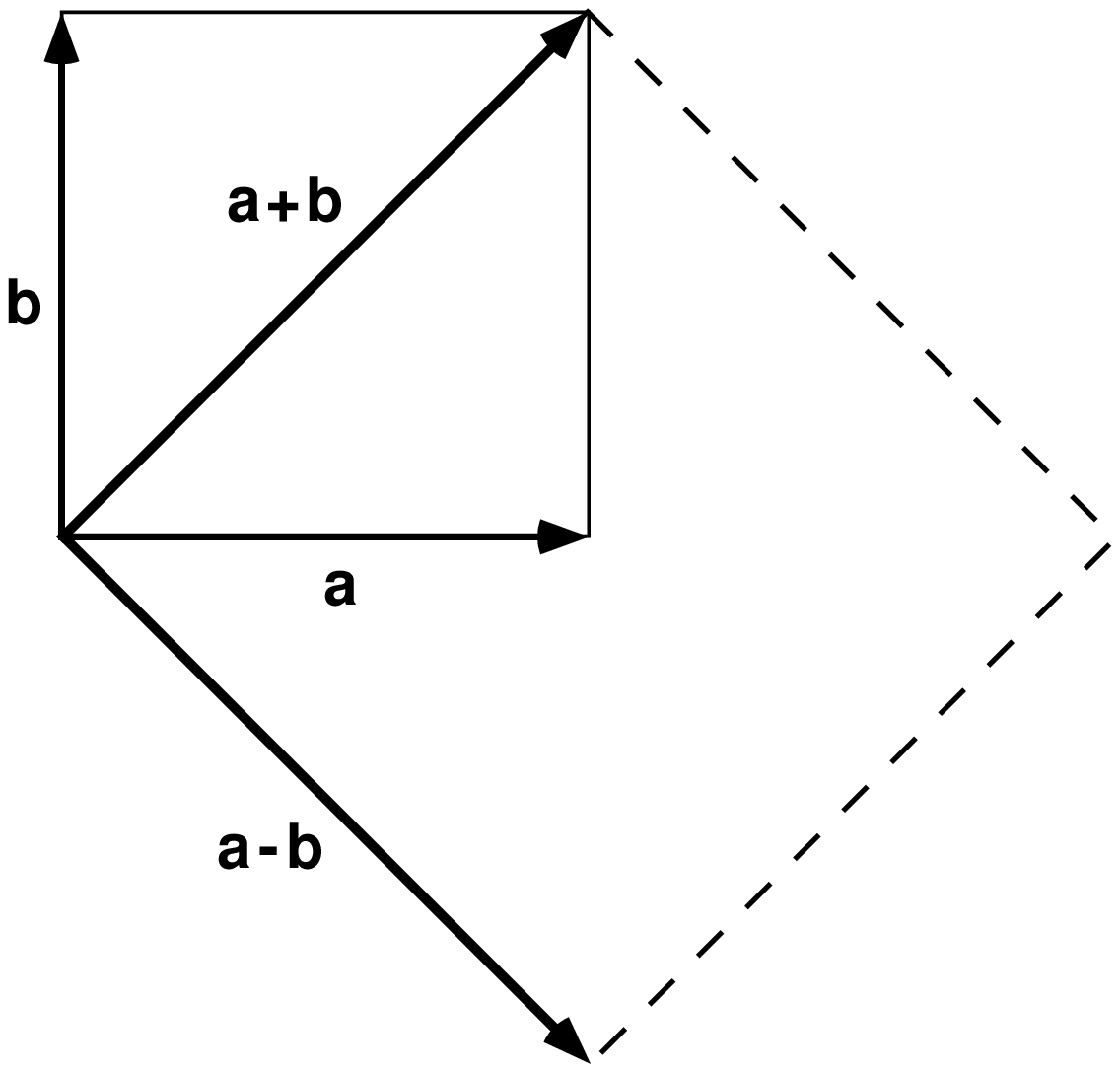}{5truecm}
\pano
This guarantees that on the oscillator modes of $X^5+X^6$ and $X^5-X^6$
the operation $\Omega P$ acts in the same way.
Open strings can wind around the $a-b$ cycle with energies 
proportional to $2\rho n^2$ and naively one might expect to find
KK states with momentum in the $a+b$ direction with energy $m^2/2\rho$.
This is however not correct, since a torus with cycles $a+b$ and $a-b$ 
is a double cover of the original torus with cycles $a$ and $b$. 
Said differently, moving the $a+b$ cycle along the $a-b$ cycle, 
we arrive at the original cycle already after half of the $a-b$ cycle. 
Thus, the relevant radius for the KK modes is not $\sqrt{2}\rho$ but 
$\rho/\sqrt{2}$. Then the cylinder and M\"obius strip amplitude for 
model III can be obtained simply by substituting
\eqn\trafo{\left(\sum_{m\in\ZZ} e^{-2\pi t{m^2\over \rho}} \right)^2 \to
   \left(\sum_{m\in\ZZ} e^{-4\pi t{m^2\over \rho}} \right)
   \left(\sum_{n\in\ZZ} e^{-4\pi t\rho n^2} \right) }
and
\eqn\trafob{\left(\sum_{m\in\ZZ} e^{-2\pi t{m^2\over \rho}} \right)
       \left(\sum_{m\in\ZZ} e^{-2\pi t{(m+{1\over 2})^2\over \rho}} \right)\to
   \left(\sum_{m\in\ZZ} e^{-4\pi t{(m+{1\over 2})^2\over \rho}} \right)
   \left(\sum_{n\in\ZZ} e^{-4\pi t\rho n^2} \right) }
in \cylres\ and \moebires. The tadpole conditions get not affected by this 
substitution and 
therefore model III also contains $32$ D8 and $\o{\rm D}$4 branes.
In the following section we discuss the massless spectrum in the open string 
sector.
\newsec{Massless spectrum in the open string sector}
Roughly speaking the open string spectrum in our models is one half of the 
massless spectrum in the associated GP model. 
Let us first discuss model I, with 16 D9 and D5 branes located at the same 
fixed point of $R$ and due to the $A_2^4$ shift with the
other 16 D9 and D5 branes located at the shifted fixed point of $R$.
Then GP obtained 
a vectormultiplet in the gauge group 
$U(8)^{(1)}_{99}\times U(8)^{(2)}_{99}\times U(8)^{(1)}_{55}\times 
U(8)^{(2)}_{55}$
and the following massless hypermultiplets 
\smno
\cl{\vbox{
\hbox{\vbox{\offinterlineskip
\def\tablespace{height2pt&\omit&&\omit&&\omit&&
 \omit&\cr}
\def\tablerule{\tablespace\noalign{\hrule}\tablespace}

\hrule\halign{&\vrule#&\strut\hskip0.2cm\hfil#\hfill\hskip0.2cm\cr
\tablespace
& $U(8)^{(1)}_{99}$ && $U(8)^{(2)}_{99}$ && $U(8)^{(1)}_{55}$ && 
$U(8)^{(2)}_{55}$  &\cr
\tablerule
& 2$\times${\bf 28} && {\bf 1} && {\bf 1} && {\bf 1} &\cr
\tablespace
& {\bf 1} && 2$\times${\bf 28} && {\bf 1} && {\bf 1} &\cr
\tablespace
& {\bf 1} && {\bf 1} && 2$\times${\bf 28} && {\bf 1} &\cr
\tablespace
& {\bf 1} && {\bf 1} && {\bf 1} && 2$\times${\bf 28} &\cr
\tablerule
& {\bf 8} && {\bf 1} && {\bf 8} && {\bf 1} &\cr
\tablespace
& {\bf 1} && {\bf 8} && {\bf 1} && {\bf 8} &\cr
\tablespace
& {\bf 8} && {\bf 1} && {\bf 1} && {\bf 8} &\cr
\tablespace
& {\bf 1} && {\bf 8} && {\bf 8} && {\bf 1} &\cr
\tablespace}\hrule}}}}
\cl{
\hbox{{\bf Table 4:}{\it ~~ GP  massless spectrum }}}
The action of the shift $A_2^4$ relates the gauge group 
$U(8)^{(1)}_{99}\times U(8)^{(1)}_{55}$ to the gauge group 
$U(8)^{(2)}_{99}\times U(8)^{(2)}_{55}$ and the matter multiplets
in subsequent rows 1 and 2, 3 and 4 and so on.  
Summarising, the massless spectrum after modding out the shift $A_2^4$
is presented in Table 5
\smno
\cl{\vbox{
\hbox{\vbox{\offinterlineskip
\def\tablespace{height2pt&\omit&&
 \omit&\cr}
\def\tablerule{\tablespace\noalign{\hrule}\tablespace}

\hrule\halign{&\vrule#&\strut\hskip0.2cm\hfil#\hfill\hskip0.2cm\cr
\tablespace
& $U(8)_{99}$ && $U(8)_{55}$  &\cr
\tablerule
& 2$\times${\bf 28} && {\bf 1} &\cr
\tablespace
& {\bf 1} && 2$\times${\bf 28} &\cr
\tablerule
& {\bf 8} && {\bf 8} &\cr
\tablespace
& {\bf 8} && {\bf 8} &\cr
\tablespace}\hrule}}}}
\cl{
\hbox{{\bf Table 5:}{\it ~~ massless spectrum after $A_2^4$ orbifold}}}
\smno
Note, that this is exactly the massless spectrum of
the orientifold of \typeb\ on a $Z_2$ orbifold of the  $SO(8)$ torus \rbackb.
Both the reduction of the rank of the gauge group and the doubling of the
states in the 95 sector is as expected from computations using the background
$B_{ij}$ field explicitly from the very beginning. \pano 
The action of $f$ relates the two gauge multiplets arising from the D9 and
D5-branes to each other, finally leading to the $N=2$ supersymmetric
massless spectrum of one vectormultiplet of  $U(8)$ and 2 hypermultiplets
in the {\bf 28} representation. Due to the shifts in the 5-6 directions, the
95 sector is  massive containing 
one hypermultiplet in the {\bf 8}$\otimes${\bf 8}$=${\bf 28}+{\bf 36} 
representation.  
As in the GP model one can move the D9- and D5- branes away from the
fixed point of $R$ leading to more general gauge groups
\eqn\eich{\prod_{I=1}^{8} U(m_I)\times \prod_{J=1}^{8} USp(n_J) }
with the constraint $\sum m_I+\sum n_J=8$.
As in the GP model one expects abelian gauge anomalies, which should
be cancelled by the Green-Schwarz mechanism \ranom. 
Except some slightly different projections of the oscillator states, the
computation of the massless spectrum in the open string sector
is absolutely identical for the models II and III. Independently of the
degree of supersymmetry in the bulk, one always gets $N=2$ supersymmetry 
on the brane. 
Thus, we have a situation which is very similar to 
the supersymmetry breaking scenario proposed by Scherk and Schwarz \rscherk. 
Quantum effects of course will mediate supersymmetry breaking from the bulk
to the brane.\pano
One can obtain more interesting spectra by putting more general Wilson lines
around the 5-6 direction. For instance, on can have two sets of 16 9-branes
with 5-6 Wilson-lines $\Theta_1$ and $\Theta_2$ respectively. Then we also 
need
two sets of 16 5-branes with the shifted Wilson lines $\Theta_1+{1\over 2R}$ 
and $\Theta_2+{1\over 2R}$. If in the other four compact dimensions
we have no further Wilson lines, we get $U(4)\times U(4)$ gauge symmetry
and matter multiplets $2\,({\bf 6,1})+ 2\,({\bf 1,6})$ arising in the
$99$ and $55$ open string sectors. If we choose   
$\Theta_2=\Theta_1+{1\over 2R}$, we can also get massless modes in the
$95$ sector.
In our case we obtain two hypermultiplets in the bi-fundamental $({\bf 4,4})$ 
representation,
so part of the $N=2$ massive spectrum becomes massless.
\newsec{Conclusions}
Technically, in this paper we have studied new kinds of orientifolds, 
namely we have arranged
asymmetric $\ZZ_2\times\ZZ_2$ orbifolds of \type, such that the two $\ZZ_2$
symmetries get exchanged by the action of world sheet parity reversal. 
It was then possible to gauge the parity reversal operation thereby
introducing 
an open string sector in the theory. The action of the former 
$\ZZ_2\times\ZZ_2$ symmetry  in the open string sectors related $Dp$ and
$D$(p-4) branes to each other.
Unaffected by the supersymmetry in the closed string sector, in all cases the 
supersymmetry in the open string sector was N=2.
We computed the remaining three one-loop contributions to the cosmological 
constant and 
found that they were all vanishing. Since there is very little known about
higher genus non-oriented string loop diagrams, we have nothing to say about
whether one might hope that the cosmological constant vanishes to higher loop 
order, as well. \pano
From a phenomenological point of view these models may be interesting, because
they combine a (possibly) vanishing cosmological constant with
non-abelian gauge symmetries living on the branes.
I would be interesting to study whether one can construct chiral models
with only $N=1$ supersymmetry on the brane in this set up. \pano
Unfortunately, the recently discussed brane-world scenario 
\refs{\rkakutye--\rshiutyeb}, where one can have
a string scale as low as 1TeV is not directly applicable. The essential 
condition there is, that one can grow a large radius transverse to the brane 
on which the gauge degrees of freedom live. In our case all radii orthogonal 
to the D(p-4)-branes are fixed at the self-dual radius and related to that 
the gauge degrees of freedom live not only on D(p-4)-branes but also on 
Dp-branes. \pano
It is known that the GP model has an F-theory dual description \rfsen, so
it would be interesting to determine to which symmetry of F-theory $f$ and
$g$ correspond to. Finally, let us make clear  that our results are not in 
contradiction to the duality of Type I 
and heterotic string theory. In the introduction we claimed that it is 
impossible for perturbative heterotic strings to be non-supersymmetric and
have vanishing cosmological constant. The orientifold models
discussed so far would however be at best dual to non-perturbative heterotic
backgrounds with also the solitonic heterotic 5-brane involved.

\bigskip\bigskip\centerline{{\bf Acknowledgments}}\pano
We would like to thank A. Karch, D. L\"ust and E. Silverstein for discussion. 
We also thank Gary Shiu for helpful comments about an earlier version
of this paper. 

\listrefs
\bye